\documentclass{article}
\usepackage[preprint]{colm2026_conference}

\usepackage{microtype}
\usepackage{hyperref}
\usepackage{url}
\usepackage{booktabs}
\usepackage{capt-of}
\usepackage{graphicx}
\usepackage{amsmath}
\usepackage{amssymb}
\usepackage{amsfonts}
\usepackage{bm}
\usepackage{multirow}
\usepackage[most]{tcolorbox}
\usepackage{xcolor}
\usepackage{colortbl}
\usepackage{lineno}

\usepackage{amsmath,amsfonts,bm}

\def\eqref#1{equation~\ref{#1}}

\def\1{\bm{1}}

\DeclareMathAlphabet{\mathsfit}{\encodingdefault}{\sfdefault}{m}{sl}
\SetMathAlphabet{\mathsfit}{bold}{\encodingdefault}{\sfdefault}{bx}{n}

\tcbuselibrary{listings}

\definecolor{darkblue}{rgb}{0, 0, 0.5}
\hypersetup{colorlinks=true, citecolor=darkblue, linkcolor=darkblue, urlcolor=darkblue}

\title{PeReGrINE: Evaluating Personalized Review Fidelity with User–Item Graph Context}

\author{
Steven Au \\
Icahn School of Medicine at Mount Sinai \\
\texttt{steven.au@mssm.edu}
\And
Baihan Lin \\
Icahn School of Medicine at Mount Sinai \\
\texttt{baihan.lin@mssm.edu}
}

\begin{document}

\ifcolmsubmission
\linenumbers
\fi

\maketitle

\begin{abstract}
We introduce PeReGrINE, a benchmark and evaluation framework for personalized review generation grounded in graph-structured user--item evidence. PeReGrINE restructures Amazon Reviews 2023 into a temporally consistent bipartite graph, where each target review is conditioned on bounded evidence from user history, item context, and neighborhood interactions under explicit temporal cutoffs. To represent persistent user preferences without conditioning directly on sparse raw histories, we compute a User Style Parameter that summarizes each user's linguistic and affective tendencies over prior reviews. This setup supports controlled comparison of four graph-derived retrieval settings: product-only, user-only, neighbor-only, and combined evidence. Beyond standard generation metrics, we introduce Dissonance Analysis, a macro-level evaluation framework that measures deviation from expected user style and product-level consensus. We also study visual evidence as an auxiliary context source and find that it can improve textual quality in some settings, while graph-derived evidence remains the main driver of personalization and consistency. Across product categories, PeReGrINE offers a reproducible way to study how evidence composition affects review fidelity, personalization, and grounding in retrieval-conditioned language models.
\end{abstract}

\section{Introduction}
\label{sec:intro}

Personalized generation is difficult to evaluate because the model has to satisfy two constraints at the same time. It has to say something correct about the item, and it has to sound like the user who is supposed to be writing. Reviews make this tension easy to see. A good review is grounded in product details, but it also reflects the reviewer's tone, sentiment, and tolerance for flaws. Standard overlap metrics only capture part of that behavior, and many existing setups do not separate which evidence source is helping with which part of the task.

Recent benchmarks have made this area easier to study. LaMP \citep{salemi2024lamp} framed personalized text generation across several tasks, and PGraphRAG \citep{au2025pgraphrag} showed that graph-based retrieval can help with personalized review writing. Still, retrieval design matters. If the retrieval query contains fragments of the target review, the search area can become narrow in a way that is hard to disentangle from leakage. We refer to our benchmark as PeReGrINE, short for Personalized Review Generation with Graph and Image-grounded Neighborhood Evidence. PeReGrINE uses a different setup. Item-side retrieval is anchored on product metadata that exists before generation, and every retrieved context is filtered by explicit temporal cutoffs.

We build PeReGrINE by restructuring Amazon Reviews 2023 \citep{hou2024bridging} into a temporally consistent bipartite graph of users, items, and reviews. Each target review is paired with bounded evidence from three sources: the user's earlier reviews, prior reviews of the same item, and product metadata. Instead of conditioning the language model directly on long and often sparse raw histories, we compute a User Style Parameter that summarizes stable linguistic and affective tendencies from a user's past reviews. This gives us a compact representation of user behavior while keeping retrieval tied to information that would be available at inference time.

The benchmark supports four retrieval settings: product-only, user-only, neighbor-only, and combined evidence. This gives a direct way to test what each source contributes. Product evidence may improve grounding. User history may improve personalization. Neighbor reviews may help with product-level consensus. To measure these trade-offs, we combine standard generation metrics with Dissonance Analysis, a macro-level evaluation framework that measures deviation from expected user style and product-level consensus. We also study visual evidence as an auxiliary context source so that the benchmark is relevant to multimodal language models as well as text-only retrieval-augmented models.

Our results are fairly consistent. Within PeReGrINE's own ablations, combined graph evidence gives the best overall balance between grounding and personalization, while visual evidence helps in some settings but does not replace graph structure. In matched reruns against LaMP-style and PGraphRAG baselines, PeReGrINE improves review-text metrics on a metadata-required subset, while the baselines remain stronger on some title and rating metrics. The contribution of PeReGrINE is mainly in data construction and evaluation, not in proposing a new training algorithm. We see it as a benchmark for studying retrieval-augmented and multimodal language models under a personalized generation setting.

Our contributions are:
\begin{itemize}
    \item We introduce PeReGrINE, a benchmark for personalized review generation built from Amazon Reviews 2023 with explicit temporal constraints and bounded graph evidence.
    \item We define a product-anchored retrieval setup that separates product grounding, user history, and neighbor consensus, and avoids using target-review fragments to drive item retrieval.
    \item We propose a compact User Style Parameter for representing persistent reviewer behavior without conditioning directly on sparse raw histories.
    \item We introduce Dissonance Analysis and use it to study how text, graph, and visual evidence affect fidelity, personalization, and product grounding.
\end{itemize}

\section{Related Work}
\label{sec:relatedworks}

\subsection{Personalized Text Generation}

Personalized generation aims to adapt model outputs to a particular user's preferences, style, or behavioral history. Early work on personalized dialogue and text generation relied on explicit user attributes or hand-written personas \citep{zhang-etal-2018-personalizing, wolf2019transfertransfotransferlearningapproach}. More recent approaches use instruction tuning, retrieval, or profile summarization to condition large language models on user histories \citep{alhafni-etal-2024-personalized, jiang-etal-2024-personallm, mysore-etal-2024-pearl}. LaMP \citep{salemi2024lamp} was especially influential because it turned personalization into a benchmark problem and made it easier to compare methods across tasks such as summarization and review synthesis.

That line of work established the importance of user context, but it does not fully answer how to combine user-specific signals with item-specific grounding. Reviews are shaped by both. A model that only sees user history may capture tone but miss the product, while a model that only sees product information may produce generic text. PeReGrINE is designed as an evaluation setting for that trade-off.

\subsection{Graph Retrieval for Personalized Reviews}

Graph-based methods provide a natural way to organize user-item interactions. In recommender systems, graph neural networks and knowledge-graph methods use local and multi-hop structure to propagate preference signals across users and items \citep{wang2019ngcf, he2020lightgcn, wang2019kgat}. Retrieval-augmented language models extend a similar idea to text generation by selecting relevant context at inference time \citep{lewis2020rag, izacard2021fid}.

For personalized reviews, this structure remains useful even when the generator has vision capabilities. Images can add product cues, but they do not remove the need to separate user evidence, item evidence, and neighborhood consensus. Existing graph-retrieval work mostly focuses on text and interaction signals, so it leaves open where visual context should enter a personalized generation pipeline: retrieval, context construction, or generation. That gap motivates PeReGrINE's benchmark design.

PGraphRAG \citep{au2025pgraphrag} is the closest prior work. PeReGrINE builds on the same general problem setting of graph-based personalized review generation, but changes the evaluation protocol in two ways. First, item-side retrieval is anchored on product metadata that is available before generation rather than on text derived from the target review. Second, all retrieved evidence is filtered by explicit temporal cutoffs. These choices make the setup closer to deployment-time retrieval, but they also change the retrieval problem itself. For that reason, published PGraphRAG numbers are not directly comparable to PeReGrINE unless both methods are rerun under the same constraints.

We want to be explicit about scope here. PeReGrINE is not meant to claim that one retrieval choice is universally better in every application. The narrower claim is that benchmark comparisons are easier to interpret when the retriever only uses information that would be available before generation. That is why we disclose the difference in retrieval protocol directly.

\subsection{Multimodal Signals in Review Settings}

Visual information has long been useful in recommendation and preference modeling. Prior work uses images and other multimodal features to improve item representations, ranking, or compatibility prediction \citep{he2016vbpr, wei2020mmgnn, luo2021clip4clip}. In generation, multimodal models have been effective for captioning, question answering, and visually grounded reasoning, but personalized language generation remains less explored.

When visual context is used for review generation, it often improves factual grounding to the item while leaving the writing style relatively generic \citep{yan2023personalized, ceylan2024words}. Our focus is narrower. We use images as auxiliary evidence inside a benchmark where the main question is how different evidence sources affect personalization and consistency. In PeReGrINE, visual evidence is useful to study, but it is not the only or primary signal.

\section{Problem Formulation}
\label{sec:problem_formulation}

We model the review ecosystem as a bipartite graph $G = (U, I, E)$, where $U$ is the set of users, $I$ is the set of items, and each edge in $E$ is a review connecting a user to an item. A review is represented as
\begin{equation}
\label{eq:review_tuple}
r = (s, t, b, \mathcal{V}, \tau),
\end{equation}
where $s$ is the scalar rating, $t$ is the review title, $b$ is the review body, $\mathcal{V}$ is optional visual evidence associated with the review, and $\tau$ is the timestamp. Each item node $i \in I$ also contains product metadata $M_i$, such as textual descriptions and catalog images.

Each benchmark instance corresponds to a target user-item pair $(u, i)$ and a gold review written at time $\tau^\star$. The model is allowed to use only evidence with timestamps earlier than $\tau^\star$. This constraint is enforced for both user history and item neighborhoods.

For a target item $i$, we define the temporally valid item neighborhood as
\begin{equation}
\label{eq:item_history}
H_i^{<\tau^\star} = \{r \mid (v, i, r) \in E,\ v \in U,\ \tau(r) < \tau^\star \}.
\end{equation}
The item profile is then
\begin{equation}
\label{eq:item_profile}
P_i = (M_i, H_i^{<\tau^\star}).
\end{equation}

For a target user $u$, the temporally valid user history is
\begin{equation}
\label{eq:user_profile}
P_u = H_u^{<\tau^\star} = \{r \mid (u, j, r) \in E,\ j \in I,\ \tau(r) < \tau^\star \}.
\end{equation}

The task is to generate a full review $\hat{R}_{\mathrm{gen}} = (\hat{s}, \hat{t}, \hat{b})$ that is grounded in the item context while reflecting the user's writing behavior.

\subsection{User Style Parameter}

Rather than conditioning directly on all of $P_u$, we compute a dense User Style Parameter
\begin{equation}
\label{eq:style_agg}
\theta_s = \mathrm{Aggregate}(P_u),
\end{equation}
which summarizes stable linguistic and affective tendencies from the user's prior reviews. This summary is used to rank evidence and to represent persistent user preferences without forcing the language model to ingest long, sparse histories.

\subsection{Operational Objective}

At the full-profile level, the task can be written as
\begin{equation}
\label{eq:formal_task}
\hat{R}_{\mathrm{gen}} = \arg\max_{R'} \Pr(R' \mid P_i, P_u).
\end{equation}
In practice, the full profiles are too large to condition on directly, so we introduce item-side and user-side retrieval functions:
\begin{equation}
\label{eq:operational_task}
\hat{R}_{\mathrm{gen}} = \arg\max_{R'} \Pr(R' \mid \mathcal{R}_i(P_i, \theta_s), \mathcal{R}_u(P_u)).
\end{equation}

One practical distinction of PeReGrINE is that $\mathcal{R}_i$ never observes the target review text. Item-side retrieval is anchored on product metadata and temporally valid item neighbors, while $\mathcal{R}_u$ draws from the user's prior reviews. This keeps the benchmark focused on inference-time evidence rather than answer-like queries.

\section{Method}
\label{sec:methodology}

\begin{figure*}[t]
    \centering
    \includegraphics[width=\textwidth]{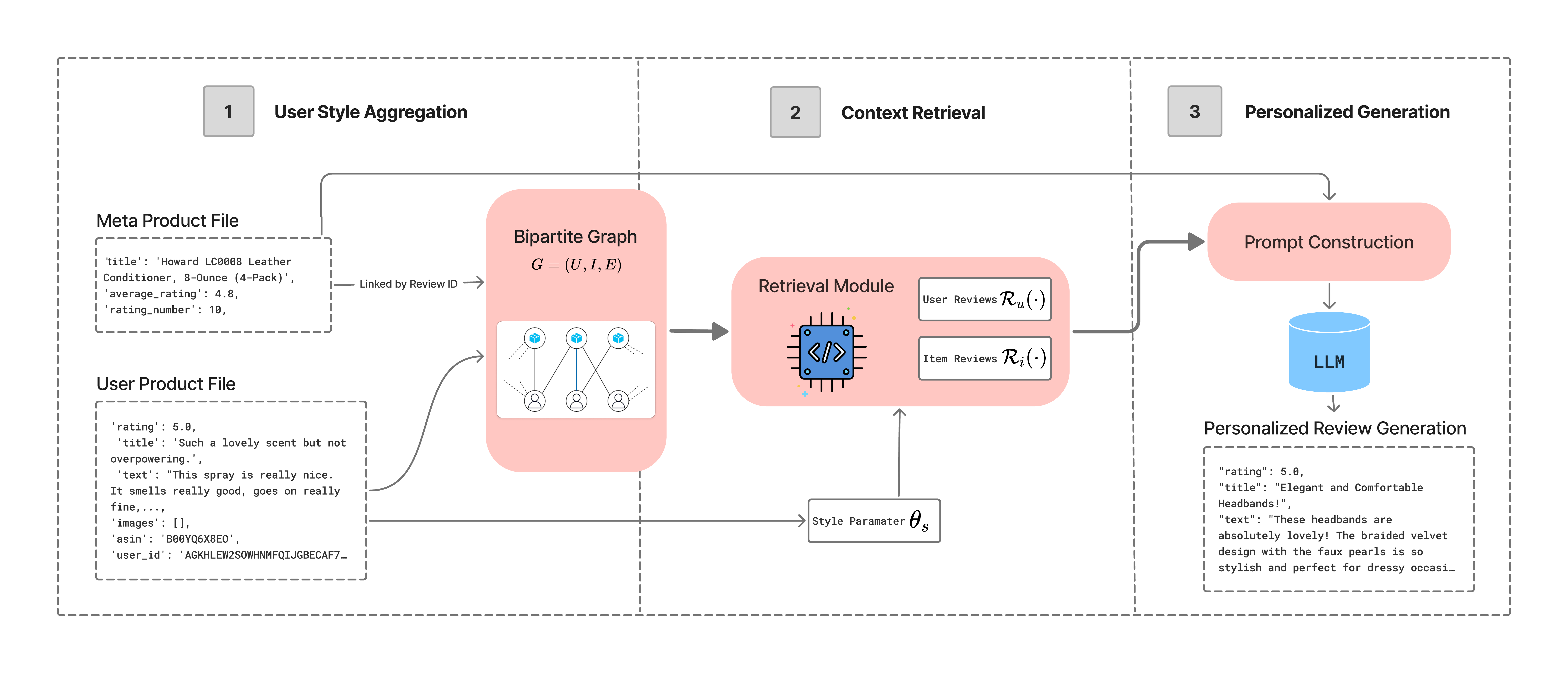}
    \caption{Overview of PeReGrINE. The system computes a user style summary, retrieves item-side and user-side evidence from the graph, and then conditions a language model on the resulting context.}
    \label{fig:framework_overview}
\end{figure*}

PeReGrINE uses a retrieval-augmented pipeline with three stages, shown in Figure~\ref{fig:framework_overview}. First, we compute a compact user style summary $\theta_s$. Second, we retrieve item evidence and user evidence from the graph under a strict temporal cutoff. Third, we prompt a language model to generate a rating, title, and review body from the selected evidence.

\subsection{User Style Aggregation}
\label{sec:style_aggregation}

To compute $\theta_s$, we extract a stylometric feature vector from each review in the user's history. The vector contains 11 explicit features: four length features (character count, word count, sentence count, and average sentence length), four sentiment features from VADER (positive, negative, neutral, and compound), and three writing-style features (punctuation density, capitalization ratio, and first-person pronoun density). We average these features over the full user history to obtain a single style vector.

This summary is not meant to replace the user's raw history in every setting. Its role is narrower. It gives the retriever a stable way to compare candidate evidence against the user's past behavior, especially when the raw history is sparse or noisy.

\subsection{Product-Anchored Item Retrieval}
\label{sec:context_retrieval_method}

The item retriever constructs a bounded item context $C_i = \mathcal{R}_i(P_i, \theta_s)$ from product metadata and temporally valid neighbor reviews. The procedure is product-anchored in the sense that the semantic query comes from $M_i$, not from any fragment of the target review.

We first gather all reviews attached to the target item that satisfy the temporal cutoff. If the neighborhood contains more than $k$ candidates, we rank each review $r \in H_i^{<\tau^\star}$ using
\begin{equation}
\label{eq:ranking_score_method}
\mathrm{score}(r) = 0.5 \times \mathrm{sim}_{\mathrm{semantic}}(r, M_i) + 0.5 \times \mathrm{sim}_{\mathrm{style}}(r, \theta_s).
\end{equation}
The semantic term is the cosine similarity between the candidate review text and the item metadata embedding. The style term is the cosine similarity between the candidate's stylometric vector and $\theta_s$.

This design keeps retrieval tied to information that would exist before generation. The retriever sees the product description, the user's historical style summary, and the graph neighborhood, but not the gold review text. That is the main methodological difference between PeReGrINE and review-text-query setups.

For visual evidence, we caption catalog images and user-provided review images with a pretrained vision-language model and store the resulting text. These captions are appended to the retrieved context after ranking. They do not determine the ranking score itself, which keeps the search process centered on graph evidence and product metadata rather than image availability.

This design also defines the current scope of the multimodal claim. PeReGrINE is multimodal at the evidence-construction and generation stages, but not in the retrieval ranking itself. Images are converted to captions and used as auxiliary context after retrieval, while the ranking function remains text- and style-based. We treat this as a deliberate simplification for benchmark design, but also as a limitation. The current benchmark does not test whether image embeddings or joint vision-text retrieval would improve item selection.

\subsection{User Retrieval}
\label{sec:user_retrieval_method}

The user retriever builds $C_u = \mathcal{R}_u(P_u)$ by selecting a small set of reviews from the target user's earlier history. Here we are not looking for product relevance. We want representative examples of the user's own voice. We therefore rank each candidate review using only the style similarity term:
\begin{equation}
\label{eq:user_ranking_score}
\mathrm{score}_{\mathrm{user}}(r) = \mathrm{sim}_{\mathrm{style}}(r, \theta_s).
\end{equation}
The top-$k_u$ reviews are passed to the language model as style evidence.

\subsection{Generation}
\label{sec:generation}

The final stage prompts a language model with the retrieved evidence. The same basic prompt template is used across product-only, user-only, neighbor-only, and combined settings; unavailable evidence blocks are simply removed. The model is asked to produce a rating, a title, and a review body in a fixed format. The full template used in the combined setting is shown in Appendix Figure~\ref{fig:prompt-template}.

\section{Experiments}
\label{sec:experiments}

\subsection{Data and Workflow}
\label{sec:dataset}

We use Amazon Reviews 2023 across 7 product categories, with summary statistics in Appendix Table~\ref{tab:dataset_stats}. Every instance satisfies a minimum user history size of $|H_u| \ge 4$ and a minimum item neighborhood size of $|H_i| \ge 3$. The training split is used only to populate the retrieval database. At evaluation time, all retrieved evidence must precede the target review timestamp.

The dataset also has a visible temporal shift. The development splits contain older reviews and a much smaller share of image-based examples than the test splits. This makes the benchmark useful for studying how retrieval settings behave under different levels of available visual context.

We also disclose an important protocol choice. PeReGrINE is not directly comparable to setups that retrieve with fragments of the target review or other answer-like text. Those queries can shrink the candidate set around the gold review. In PeReGrINE, item retrieval is anchored on product metadata and prior graph evidence only. We make this choice to reduce leakage risk and to keep the evaluation closer to what would be available before generation. It may also make retrieval harder, so the distinction should be kept in mind when comparing against prior work.

We run four experimental stages:
\begin{enumerate}
    \item A text-only ablation on the All Beauty subset using Qwen2.5-3B-Instruct, comparing product-only, user-only, neighbor-only, and combined evidence.
    \item A text-only model comparison on the combined setting for All Beauty.
    \item A multimodal ablation on All Beauty using Qwen3-VL-8B-Instruct.
    \item A final category-level run with the combined multimodal setting on the full 500-instance test sets for all 7 categories.
\end{enumerate}

\subsection{Evidence Settings}
\label{sec:exp_conditions}

To isolate the effect of each evidence source, we evaluate four retrieval settings:
\begin{enumerate}
    \item Product: item metadata $M_i$ only.
    \item User: user history $H_u$ only.
    \item Neighbor: prior reviews of the target item $H_i$ only.
    \item Both: product metadata, user history, and item neighbors together.
\end{enumerate}

\subsection{Matched Baselines}
\label{sec:baseline_protocol}

For a fair comparison, we rerun LaMP-style prompting, PGraphRAG, and PeReGrINE on the same metadata-required subset of 100 All Beauty development examples, using the same temporal cutoffs, generator path, prompt budget, and output format. We also test a joint PeReGrINE variant that writes one full review and then projects that output to title, text, and rating evaluation.

We do not treat published results from prior work as directly comparable because the retrieval protocol is different. PeReGrINE anchors item retrieval on product metadata and prior graph evidence only, so a matched rerun is the appropriate comparison. All reruns passed the minimum-$k$ constraints, user and neighbor invariants, parsing checks, and non-empty-query checks. One caveat is that the OpenAI Responses path used for these reruns did not support explicit seed control.

\begin{table}[t]
\centering
\caption{Matched baseline comparison on the metadata-required 100-example All Beauty development subset. Higher is better for ROUGE-L, BERTScore-F1, and METEOR. Lower is better for MAE and RMSE.}
\label{tab:baseline_stub}
\small
\setlength{\tabcolsep}{3pt}
\begin{tabular}{lcccccccc}
\toprule
Method & Text R-L & Text B-F1 & Text M & Title R-L & Title B-F1 & Title M & MAE & RMSE \\
\midrule
LaMP & 0.1452 & 0.7573 & 0.1180 & \textbf{0.1593} & \textbf{0.7477} & \textbf{0.1468} & 0.30 & 0.6164 \\
PGraphRAG & 0.1475 & 0.7551 & 0.1259 & 0.1459 & 0.7365 & 0.1303 & \textbf{0.29} & \textbf{0.5916} \\
PeReGrINE & \textbf{0.1595} & \textbf{0.7636} & 0.1443 & 0.1474 & 0.7400 & 0.1343 & 0.33 & 0.6708 \\
PeReGrINE-joint & 0.1277 & 0.7543 & \textbf{0.1665} & 0.0465 & 0.7092 & 0.0304 & 0.94 & 1.5427 \\
\bottomrule
\end{tabular}
\end{table}

Table~\ref{tab:baseline_stub} adds the main comparison missing from the earlier draft. On this matched subset, PeReGrINE gives the strongest review-text ROUGE-L and BERTScore-F1, while LaMP gives the strongest title metrics and PGraphRAG gives the lowest rating error. The joint variant has the highest review-text METEOR, but that appears to come from longer generations rather than better task fidelity. This pattern fits the retrieval designs: PeReGrINE is optimized for grounded review generation, while rating prediction is less directly targeted than in query styles that place more weight on answer-like sentiment cues.

The joint variant performs substantially worse, especially on title and rating metrics. This is mainly a mode mismatch rather than a retrieval failure. In the split setting, each task is prompted separately. In the joint setting, one review is reused for all three evaluations. On this subset, the joint model predicts a 5-star rating in 85 of 100 cases, compared with 54 gold 5-star ratings, and produces much longer reviews on average than the split version. The paired significance tests also show strong split-versus-joint differences. For example, joint versus split PeReGrINE yields $p=1.7 \times 10^{-7}$ for title ROUGE-L and $p=3.56 \times 10^{-5}$ for absolute rating error.

\subsection{Evaluation}
\label{sec:exp_metrics}

We use two groups of metrics. The first group measures generation quality against the gold review. For review text and title generation, we report ROUGE-L, BLEU, METEOR, and BERTScore-F1. For rating prediction, we report exact-match accuracy, MAE, and RMSE. We also report title-text consistency to measure whether the generated title matches the generated review body.

The second group is Dissonance Analysis. These macro-level scores measure deviation from expected user and product behavior, with lower values indicating better alignment. User Dissonance measures how far the output moves away from the user's historical style, sentiment, length, and rating behavior. Product Dissonance measures how far the generated review moves away from the product's consensus cluster. Sentiment Dissonance measures disagreement among the generated text, the predicted rating, and the gold review. Appendix~\ref{sec:appendix_dissonance} provides the operational formulas used for these three scores.

\subsection{Results}
\label{sec:results}

We first analyze the evidence ablations on All Beauty, then compare models in the combined text-only setting, then test multimodal evidence, and finally report category-level results.

\subsubsection{Text-Only Evidence Ablation}
\label{sec:results_context_ablation}

Table~\ref{tab:micro_metrics} and Table~\ref{tab:macro_metrics} show a clear trade-off across evidence sources. Product-only evidence yields the strongest grounding, with the best Product Dissonance and the strongest title-text consistency. User-only evidence gives the best personalization signal, with the lowest User Dissonance and the highest rating accuracy. The combined setting gives the best overall text similarity and the most balanced behavior across the metric groups.

\begin{table}[t]
\centering
\caption{Micro metrics for the text-only evidence ablation on All Beauty using Qwen2.5-3B-Instruct. Best values are in bold.}
\label{tab:micro_metrics}
\small
\setlength{\tabcolsep}{5pt}
\begin{tabular}{llcccc}
\toprule
Task & Metric & Product & User & Neighbor & Both \\
\midrule
\multirow{4}{*}{Text}
  & ROUGE-L\,($\uparrow$)   & 0.110 & 0.114 & 0.115 & \textbf{0.123} \\
  & BLEU\,($\uparrow$)      & 0.008 & 0.011 & 0.005 & \textbf{0.016} \\
  & METEOR\,($\uparrow$)    & \textbf{0.181} & 0.160 & 0.163 & 0.173 \\
  & BERT-F1\,($\uparrow$)   & 0.742 & 0.743 & 0.746 & \textbf{0.752} \\
\addlinespace
\multirow{2}{*}{Title}
  & ROUGE-L\,($\uparrow$)   & 0.058 & 0.032 & 0.044 & \textbf{0.060} \\
  & BERT-F1\,($\uparrow$)   & \textbf{0.707} & 0.699 & 0.697 & \textbf{0.707} \\
\addlinespace
Title-Text
  & Consistency\,($\uparrow$) & \textbf{0.592} & 0.334 & 0.306 & 0.368 \\
\addlinespace
\multirow{3}{*}{Rating}
  & Accuracy\,($\uparrow$)  & 0.261 & \textbf{0.421} & 0.396 & 0.371 \\
  & MAE\,($\downarrow$)     & 0.903 & 0.817 & 0.753 & \textbf{0.705} \\
  & RMSE\,($\downarrow$)    & 1.154 & 1.321 & 1.195 & \textbf{1.153} \\
\bottomrule
\end{tabular}
\end{table}

\begin{table}[t]
\centering
\caption{Dissonance metrics for the text-only evidence ablation on All Beauty using Qwen2.5-3B-Instruct. Lower is better.}
\label{tab:macro_metrics}
\small
\begin{tabular}{lcccc}
\toprule
Metric $\downarrow$ & Product & User & Neighbor & Both \\
\midrule
User Dissonance      & 0.274 & \textbf{0.199} & 0.257 & 0.251 \\
Product Dissonance   & 0.358 & 0.503 & 0.400 & \textbf{0.399} \\
Sentiment Dissonance & 0.164 & 0.164 & 0.169 & \textbf{0.162} \\
\bottomrule
\end{tabular}
\end{table}

These results help clarify what the benchmark is measuring. User evidence and product evidence do not solve the same problem. One mainly helps the model sound like the reviewer, and the other mainly helps it stay close to the product. The combined setting works best because it exposes both signals at once.

\subsubsection{Text-Only Model Comparison}
\label{sec:results_model_comparison}

We next compare several text-only models on the combined All Beauty setting. The full tables are reported in Appendix Table~\ref{tab:model_comp_micro} and Appendix Table~\ref{tab:model_comp_macro}. LLaMA-3.1-8B-Instruct gives the strongest lexical overlap on the review body, Claude-4.5-Haiku gives the strongest BERTScore and title overlap, and GPT-5-nano is best on rating prediction and title-text consistency. This comparison shows that the benchmark is not only measuring one kind of fluency. Different models make different trade-offs among lexical similarity, rating behavior, and macro-level alignment.

\subsubsection{Multimodal Ablation}
\label{sec:results_multimodal}

We next ask whether adding images changes the ranking of evidence sources established in the text-only setting. We repeat the four-setting ablation with Qwen3-VL-8B-Instruct on the same All Beauty slice. The answer is mostly no. Combined evidence still gives the strongest overall performance, user-only still gives the lowest User Dissonance, and product-only remains competitive on grounding-oriented metrics. Images help mostly at the margin, improving several text metrics and slightly reducing product-level dissonance, but they do not overturn the structure already seen in the text-only case.

\begin{table}[t]
\centering
\caption{Micro metrics for the multimodal evidence ablation on All Beauty using Qwen3-VL-8B-Instruct.}
\label{tab:vision_micro_metrics}
\small
\setlength{\tabcolsep}{5pt}
\begin{tabular}{llcccc}
\toprule
Task & Metric & Product & User & Neighbor & Both \\
\midrule
\multirow{4}{*}{Text}
  & ROUGE-L\,($\uparrow$)     & 0.127 & 0.120 & 0.124 & \textbf{0.131} \\
  & BLEU\,($\uparrow$)        & 0.012 & 0.009 & 0.008 & \textbf{0.014} \\
  & METEOR\,($\uparrow$)      & 0.183 & 0.135 & 0.164 & \textbf{0.192} \\
  & BERT-F1\,($\uparrow$)     & 0.752 & 0.744 & 0.753 & \textbf{0.759} \\
\addlinespace
\multirow{2}{*}{Title}
  & ROUGE-L\,($\uparrow$)     & 0.072 & 0.049 & 0.040 & \textbf{0.059} \\
  & BERT-F1\,($\uparrow$)     & 0.717 & 0.701 & 0.712 & \textbf{0.719} \\
\addlinespace
Title-Text
  & Consistency\,($\uparrow$) & \textbf{0.541} & 0.336 & 0.443 & 0.469 \\
\addlinespace
\multirow{3}{*}{Rating}
  & Accuracy\,($\uparrow$)    & 0.517 & 0.538 & 0.558 & \textbf{0.588} \\
  & MAE\,($\downarrow$)       & 0.746 & 0.854 & 0.752 & \textbf{0.692} \\
  & RMSE\,($\downarrow$)      & \textbf{1.243} & 1.422 & 1.301 & 1.248 \\
\bottomrule
\end{tabular}
\end{table}

\begin{table}[t]
\centering
\caption{Dissonance metrics for the multimodal evidence ablation on All Beauty. Lower is better.}
\label{tab:vision_macro_metrics}
\small
\begin{tabular}{lcccc}
\toprule
Metric $\downarrow$ & Product & User & Neighbor & Both \\
\midrule
User Dissonance      & 0.253 & \textbf{0.197} & 0.253 & 0.252 \\
Product Dissonance   & \textbf{0.374} & 0.501 & 0.379 & \textbf{0.374} \\
Sentiment Dissonance & \textbf{0.150} & 0.158 & 0.161 & 0.159 \\
\bottomrule
\end{tabular}
\end{table}

Appendix Figure~\ref{fig:sig4panel} points in the same direction. Images improve several text-body and title metrics, but the improvement in rating accuracy is limited. Visual evidence helps, especially for product grounding and descriptive quality, but it is still auxiliary to the graph-derived evidence. That is the framing we carry into the full-category run.

\subsubsection{Category-Level Results}
\label{sec:results_categorical}

After confirming on All Beauty that images help but do not replace graph structure, we run the combined multimodal configuration on the full 500-instance test sets for all 7 categories. Appendix Figure~\ref{fig:category-comparison} and the category-wise appendix tables show large differences across domains. All Beauty is the easiest category in terms of text overlap and product-level consistency. Sports and Toys \& Games show stronger rating predictability but less stable text quality. This suggests that category structure still matters more than the mere presence of visual context: some domains have tight product consensus and formulaic wording, while others allow more narrative variation.

\section{Conclusion}
\label{sec:conclusion}

PeReGrINE turns Amazon Reviews 2023 into a benchmark for personalized review generation with explicit temporal constraints and product-anchored graph retrieval. The benchmark separates product evidence, user history, and neighbor consensus, and represents reviewer behavior with a compact style parameter rather than relying only on long raw histories.

Across the evaluated settings, the clearest result is that graph-derived evidence is the main driver of personalization and consistency. Product evidence improves grounding, user evidence improves stylistic alignment, and the combined setting gives the best overall balance. Visual evidence can improve text quality and product alignment in some cases, but it acts as an auxiliary signal rather than the main source of personalization.
We view PeReGrINE as a controlled evaluation setting rather than a final system. Its main purpose is to make it easier to study how retrieval design choices affect grounding, leakage risk, and behavioral fidelity in personalized generation. In that sense, the paper is mainly about benchmark design, evaluation protocol, retrieval-augmented language models, and multimodal evidence rather than about introducing a new learning algorithm. The next sections state the main future directions, limitations, and disclosure details directly in the main paper.

\section*{Future Work}

The clearest next step is to make retrieval itself multimodal. The current benchmark adds image evidence after ranking, but the ranking function still relies on text and stylometric similarity. A stronger follow-up would test image embeddings, joint vision-text retrieval, and visually similar-item retrieval so that visual evidence affects candidate selection directly rather than only the final prompt.

Another useful extension is to broaden the product-side evidence. Product videos, video-derived features, frame-level descriptions, and explicit visual attribute extraction from catalog media would make the benchmark more relevant to multimodal language models used in realistic shopping settings. Richer review-side visual evidence would also help test how user-posted media interacts with product grounding.

Two evaluation extensions are also still open. The first is feature-group ablations for the User Style Parameter so that length, sentiment, and writing-style cues can be tested separately. The second is component ablations for the Dissonance metrics so it is clearer which terms drive the final scores.

\section*{Limitations}

PeReGrINE is designed as a controlled benchmark, not as a final multimodal retrieval system. The retrieval ranker is not yet multimodal, so the current setup cannot test whether direct visual retrieval would improve item selection. The benchmark also favors denser graph neighborhoods over sparse cold-start cases, which improves experimental control but shifts the setting away from the hardest recommendation regimes.

The User Style Parameter is intentionally compact and interpretable, but it is still only a proxy for reviewer behavior. It does not fully capture richer lexical, syntactic, or discourse-level variation, and this submission does not include a feature-group ablation to separate the contribution of each part of the style vector.

The Dissonance metrics should also be read as heuristic summaries rather than exact measures. They are useful for comparing behavioral drift across settings, but the current submission does not include component ablations that isolate the effect of each term. Appendix~\ref{sec:appendix_style} and Appendix~\ref{sec:appendix_metric_ablation} provide the underlying definitions and the current scope of these choices.

\section*{AI Use Disclosure}

Generative AI tools were used only for grammar and formatting suggestions during manuscript preparation and for limited help with chart generation from experimental outputs. All charts were checked manually. Diagrams and illustrations were created manually by the authors. All experimental design decisions, analyses, and final wording choices remained under author control.

\bibliography{main}
\bibliographystyle{colm2026_conference}

\appendix
\definecolor{goldbg}{RGB}{240,248,240}
\definecolor{textbg}{RGB}{250,240,240}
\definecolor{multibg}{RGB}{240,245,255}

\section{Data Processing Pipeline}
\label{sec:appendix_data}

To construct PeReGrINE, we processed Amazon Reviews 2023 \citep{hou2024bridging} with a filtering and indexing pipeline designed to preserve temporal integrity while remaining practical for large product categories.

\subsection{Pre-Processing and Graph Construction}

The raw dataset is sparse and noisy. We retained only reviews posted after January 1, 2016, removed duplicates, and enforced minimum interaction counts for both items and users. In the final benchmark construction, we required at least three prior item-neighbor reviews and at least four prior user reviews so that each target instance had enough graph evidence for retrieval.

To preserve temporal generalization, users were sorted by their most recent review timestamp and partitioned into train, development, and test splits. For every evaluation instance with gold timestamp $t_{\mathrm{gold}}$, any user-history review or item-neighbor review with timestamp $t \ge t_{\mathrm{gold}}$ was discarded. This cutoff is applied before retrieval and before prompt construction.

\subsection{Scalable Neighbor Retrieval}

A practical difficulty in graph-based benchmarking is the cost of validating temporally admissible neighbors. A naive linear scan becomes expensive for large categories. We therefore indexed each product by timestamp-sorted review lists and used binary search to locate valid neighbors. This reduced the cost of temporal validation substantially and made full-category preprocessing feasible without heavy subsampling.

\subsection{Selection Trade-Offs}

The benchmark uses different selection priorities at different stages. Early ablations on All Beauty emphasized graph structure and stable comparisons across retrieval settings. The broader category evaluation emphasized dense graph neighborhoods and sufficient image availability so that multimodal effects were observable. This choice improves controllability, but it also biases the benchmark away from the most sparse cold-start cases. We therefore treat sparsity robustness as an open question rather than a solved one.

\begin{table*}[t]
\centering
\caption{Dataset statistics. All instances satisfy $|H_u| \ge 4$ and $|H_i| \ge 3$.}
\label{tab:dataset_stats}
\small
\setlength{\tabcolsep}{4pt}
\renewcommand{\arraystretch}{0.95}
\begin{tabular}{@{}llrrrc@{}}
\toprule
\textbf{Category} & \textbf{Split} & \textbf{\# Inst.} & \textbf{Mean $|H_u|$} & \textbf{Mean $|H_i|$} & \textbf{Img. \%} \\
\midrule
\multirow{2}{*}{All Beauty}
    & Dev  & 218   & 6.07 & 6.46   & 17.0\% \\
    & Test & 399   & 9.72 & 5.81   & 21.6\% \\
\midrule
\multirow{2}{*}{Baby}
    & Dev  & 500   & 5.53 & 15.53  & 0.6\% \\
    & Test & 500   & 6.53 & 185.92 & 13.0\% \\
\midrule
\multirow{2}{*}{Arts \& Crafts}
    & Dev  & 500   & 5.53 & 12.20  & 0.8\% \\
    & Test & 500   & 7.39 & 69.13  & 12.6\% \\
\midrule
\multirow{2}{*}{Toys \& Games}
    & Dev  & 500   & 5.55 & 11.31  & 0.4\% \\
    & Test & 500   & 6.73 & 97.92  & 10.0\% \\
\midrule
\multirow{2}{*}{Pet Supplies}
    & Dev  & 500   & 5.26 & 18.77  & 1.0\% \\
    & Test & 500   & 6.76 & 343.47 & 9.0\% \\
\midrule
\multirow{2}{*}{Sports}
    & Dev  & 500   & 5.32 & 20.69  & 0.8\% \\
    & Test & 500   & 6.42 & 112.89 & 7.0\% \\
\midrule
\multirow{2}{*}{Beauty (Pers.)}
    & Dev  & 500   & 5.32 & 17.51  & 1.6\% \\
    \cline{2-6}
    & Test & 500   & 6.51 & 228.81 & 9.8\% \\
\midrule
\multirow{2}{*}{\textbf{Total}}
    & Dev  & 3,218 & -    & -      & 2.0\% \\
    & Test & 3,399 & -    & -      & 11.6\% \\
\bottomrule
\end{tabular}
\end{table*}

\section{Data Selection and Sampling Rationale}
\label{sec:appendix_data_rationale}

PeReGrINE moves from very large raw interaction logs to a smaller, denser benchmark. This was done to study personalization and grounding under controlled retrieval conditions rather than to approximate the raw distribution directly.

\subsection{Graph Connectivity}

Sparse nodes are common in recommender data, but they provide weak evidence for a benchmark whose goal is to compare user history, item context, and neighbor consensus under fixed prompt budgets. By enforcing minimum history and neighborhood sizes, the benchmark ensures that the four retrieval settings are all well-defined and that multimodal evidence is not swamped by missing context.

\begin{table*}[t]
\centering
\caption{Comparison of the raw data distribution and the denser PeReGrINE benchmark subset. The benchmark intentionally shifts toward higher-degree items so that graph-conditioned generation is well defined.}
\label{tab:density_bias}
\begin{tabular}{@{}lrrrr@{}}
\toprule
& \multicolumn{2}{c}{Raw Distribution} & \multicolumn{2}{c}{PeReGrINE Test Set} \\
\cmidrule(lr){2-3} \cmidrule(lr){4-5}
Category & Avg. Degree & Sparsity & Avg. Degree & Density Shift \\
\midrule
All Beauty             & 6.23  & High & 5.81   & -6.7\% \\
Baby Products          & 27.56 & Med  & 185.92 & +574\% \\
Arts, Crafts \& Sewing & 11.23 & Med  & 69.13  & +515\% \\
Pet Supplies           & 34.10 & Low  & 343.47 & +907\% \\
Toys \& Games          & 18.30 & Med  & 97.92  & +435\% \\
Sports \& Outdoors     & 12.25 & Med  & 112.89 & +821\% \\
\bottomrule
\end{tabular}
\end{table*}

\subsection{Cross-Category Generalization}

The final benchmark emphasizes breadth across categories rather than exhaustive depth inside a single one. This makes it possible to compare retrieval behavior across domains with very different narrative styles, rating patterns, and graph densities. At the same time, it means the benchmark is not intended to represent the raw long-tail distribution exactly.

\section{Prompt and Summary Figures}
\label{sec:appendix_prompt_figures}

\begin{figure*}[t]
  \centering
  \begin{minipage}{0.82\textwidth}
    \begin{tcblisting}{
      width=\linewidth,
      listing only,
      enhanced jigsaw,
      left=1mm, right=1mm, top=1mm, bottom=1mm,
      listing options={
        basicstyle=\ttfamily\footnotesize,
        breaklines=true,
        breakatwhitespace=false,
        columns=fullflexible,
        keepspaces=true,
        showstringspaces=false
      }
    }
Given the following product information and review evidence:

=== PRODUCT INFORMATION ===
[Formatted product metadata from C_i]

=== USER HISTORY ===
[Representative reviews from C_u]
Review 1: Rating: [r]/5, Title: [t], Text: [b]
Review 2: Rating: [r]/5, Title: [t], Text: [b]

=== ITEM NEIGHBOR REVIEWS ===
[Retrieved reviews for the same item]
Review 1: Rating: [r]/5, Title: [t], Text: [b]
Review 2: Rating: [r]/5, Title: [t], Text: [b]

Generate a complete review for this product from this user.
Include a rating from 1 to 5, a title, and review text.
Return only:
Rating: ...
Title: ...
Review: ...
    \end{tcblisting}
  \end{minipage}
  \caption{Prompt template for the combined-evidence setting. Other retrieval settings use the same template with the relevant evidence blocks removed.}
  \label{fig:prompt-template}
\end{figure*}

\begin{figure*}[t]
\centering
\begin{minipage}[t]{0.49\textwidth}
\centering
\includegraphics[width=\linewidth]{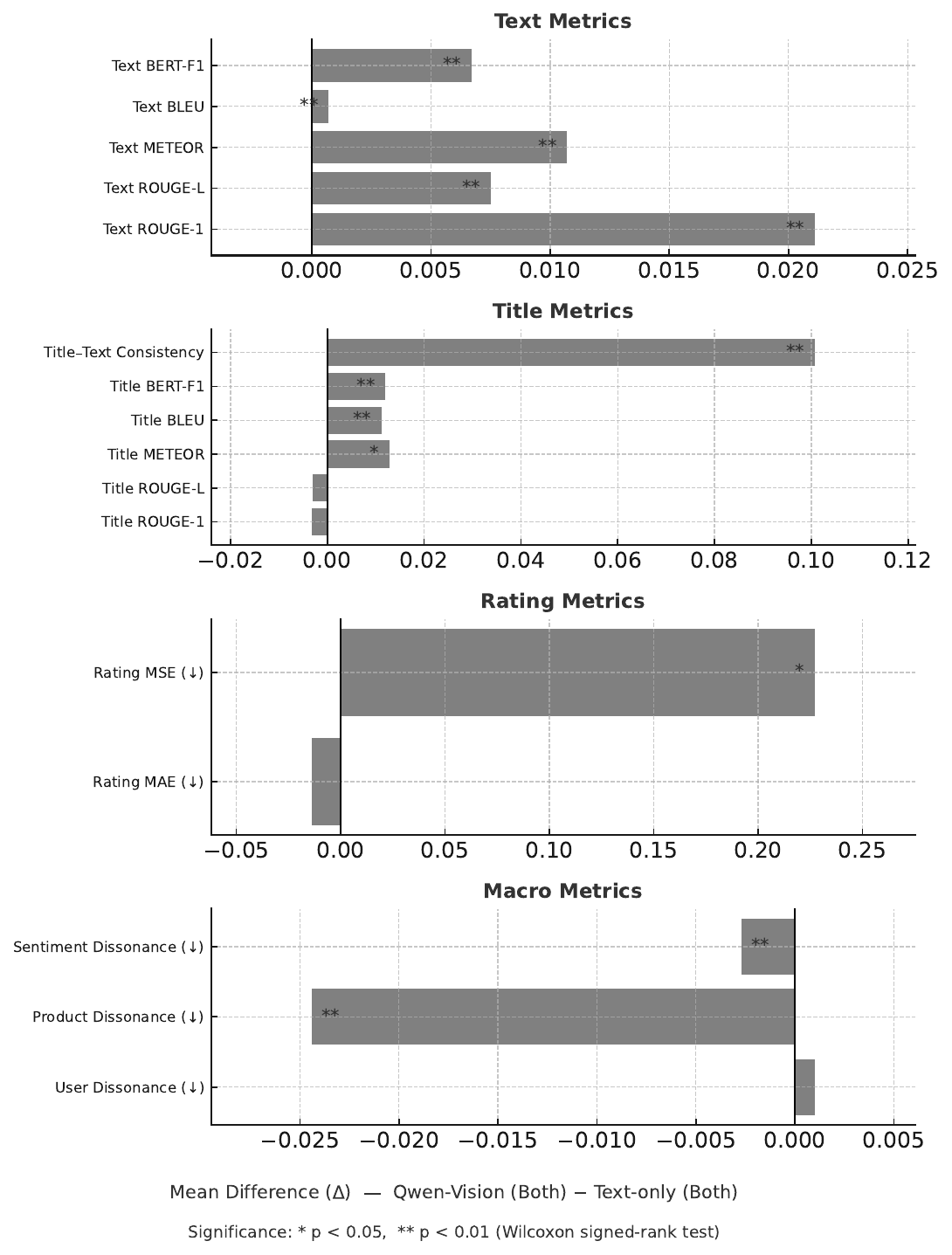}
\captionof{figure}{Per-metric mean differences between the multimodal combined setting and a text-only combined setting, grouped by task. Positive values indicate improvements from adding images. Asterisks mark significant differences under a Wilcoxon signed-rank test.}
\label{fig:sig4panel}
\end{minipage}\hfill
\begin{minipage}[t]{0.49\textwidth}
\centering
\includegraphics[width=\linewidth]{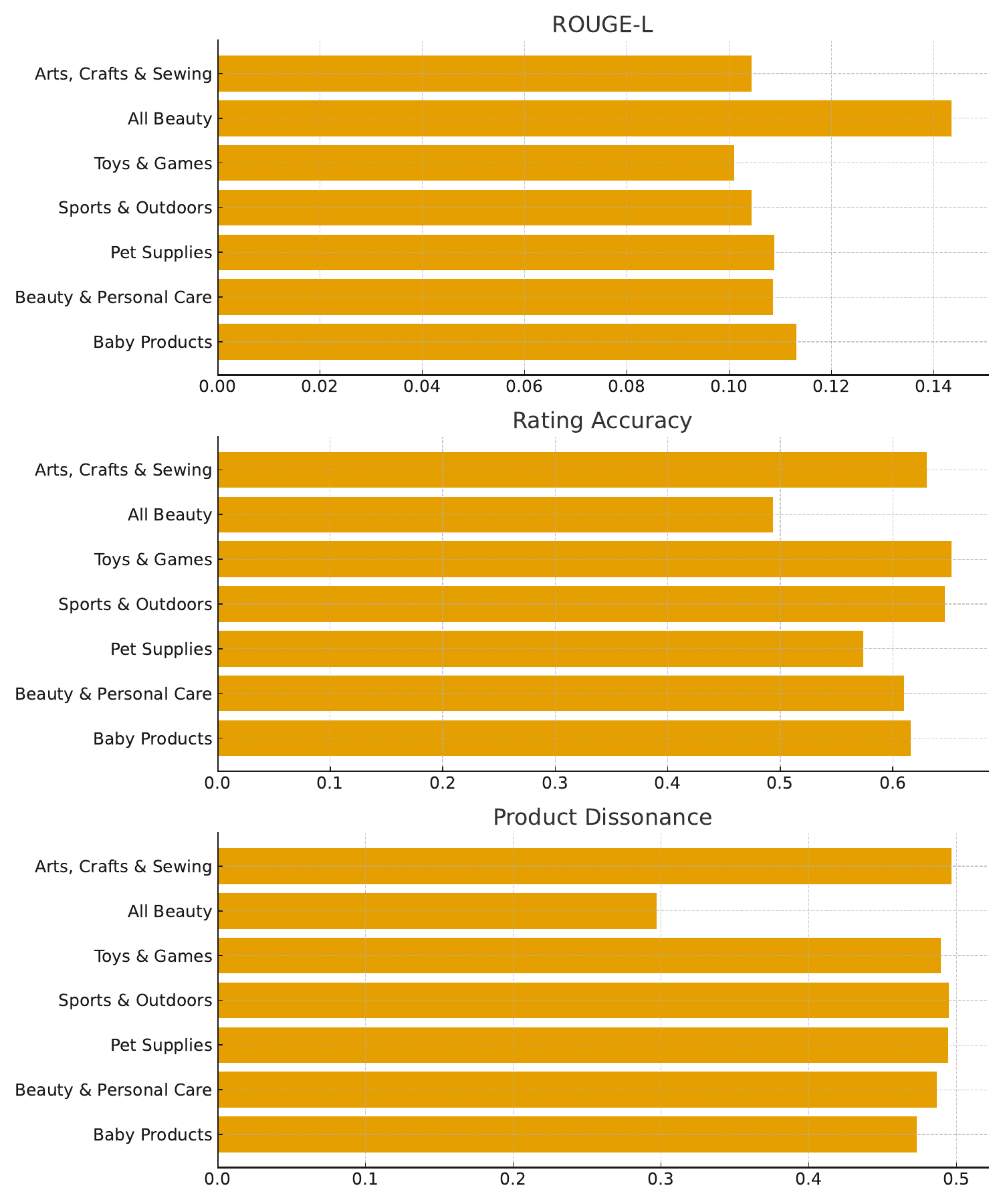}
\captionof{figure}{Category-wise performance for the combined multimodal setting on the full test sets. Top: text generation quality (ROUGE-L). Middle: rating prediction accuracy. Bottom: product dissonance.}
\label{fig:category-comparison}
\end{minipage}
\end{figure*}

\section{Graph-Based Retrieval Paradigms}
\label{sec:appendix_graph_mechanics}

\begin{figure*}[t]
    \centering
    \includegraphics[width=0.85\textwidth]{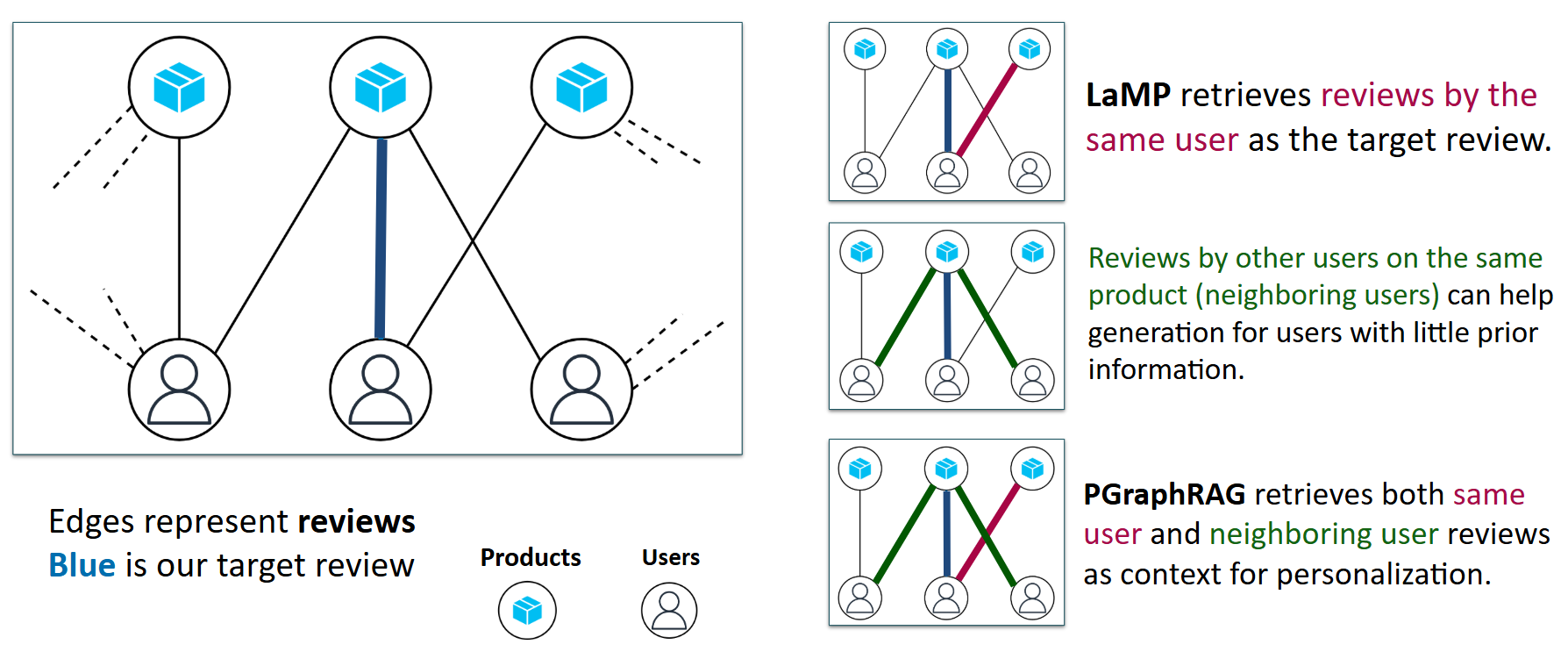}
    \caption{Retrieval paradigms supported by the bipartite formulation. User-history retrieval corresponds to the red edges, item-neighbor retrieval corresponds to the green edges, and the combined setting uses both.}
    \label{fig:retrieval_strategies}
\end{figure*}

The bipartite graph view makes it easy to separate user-side and item-side evidence while keeping both inside a single retrieval framework. Figure~\ref{fig:retrieval_strategies} illustrates this relation.

This figure is not intended as a claim of architectural novelty by itself. Its purpose is to clarify how PeReGrINE organizes evidence sources within one benchmark. In particular, the framework lets us compare user-history retrieval, neighbor retrieval, and combined retrieval under the same temporal constraints and with the same generator.

PGraphRAG remains the closest prior work. The main difference is that PeReGrINE treats product metadata as the anchor for item-side retrieval and avoids using target-review fragments as part of the retrieval query. This changes the retrieval problem and is one reason matched reruns are necessary for fair comparison.

\section{Category-Level Analysis}
\label{sec:appendix_category_analysis}

Aggregate metrics hide large category-level differences. The tables below summarize how text generation, title generation, rating prediction, and Dissonance Analysis vary across product domains.

\begin{table*}[t]
\centering
\small
\caption{Category-wise review-text generation results. Higher values indicate stronger alignment with reference reviews.}
\label{tab:category_text_metrics}
\begin{tabular}{lcccc}
\toprule
Category & ROUGE-L & BLEU & METEOR & BERTScore-F1 \\
\midrule
Baby Products           & 0.1131 & 0.0066 & 0.1863 & 0.7485 \\
Beauty \& Personal Care & 0.1086 & 0.0087 & 0.1736 & 0.7414 \\
Pet Supplies            & 0.1088 & 0.0045 & 0.1764 & 0.7430 \\
Sports \& Outdoors      & 0.1043 & 0.0072 & 0.1725 & 0.7426 \\
Toys \& Games           & 0.1009 & 0.0076 & 0.1810 & 0.7401 \\
All Beauty              & 0.1434 & 0.0136 & 0.2119 & 0.7689 \\
Arts, Crafts \& Sewing  & 0.1044 & 0.0089 & 0.1804 & 0.7411 \\
\bottomrule
\end{tabular}
\end{table*}

\begin{table*}[t]
\centering
\small
\caption{Category-wise title generation results.}
\label{tab:category_title_metrics}
\begin{tabular}{lccc}
\toprule
Category & Title ROUGE-L & Title BERTScore-F1 & Title-Text Consistency \\
\midrule
Baby Products           & 0.0626 & 0.7099 & 0.4286 \\
Beauty \& Personal Care & 0.0600 & 0.7066 & 0.4217 \\
Pet Supplies            & 0.0783 & 0.7146 & 0.4810 \\
Sports \& Outdoors      & 0.0688 & 0.7175 & 0.4413 \\
Toys \& Games           & 0.0657 & 0.7140 & 0.4601 \\
All Beauty              & 0.0640 & 0.7199 & 0.4892 \\
Arts, Crafts \& Sewing  & 0.0544 & 0.7131 & 0.4578 \\
\bottomrule
\end{tabular}
\end{table*}

\begin{table*}[t]
\centering
\small
\caption{Category-wise rating prediction results. Lower is better for MAE and RMSE.}
\label{tab:category_rating_metrics}
\begin{tabular}{lccc}
\toprule
Category & Accuracy & MAE & RMSE \\
\midrule
Baby Products           & 0.6160 & 0.8370 & 1.5190 \\
Beauty \& Personal Care & 0.6100 & 0.9660 & 1.7181 \\
Pet Supplies            & 0.5740 & 1.0360 & 1.7672 \\
Sports \& Outdoors      & 0.6460 & 0.6990 & 1.3544 \\
Toys \& Games           & 0.6520 & 0.8050 & 1.5305 \\
All Beauty              & 0.4937 & 0.7657 & 1.2163 \\
Arts, Crafts \& Sewing  & 0.6300 & 0.8080 & 1.5205 \\
\bottomrule
\end{tabular}
\end{table*}

\begin{table*}[t]
\centering
\small
\caption{Category-wise Dissonance metrics. Lower values indicate better alignment.}
\label{tab:category_dissonance_metrics}
\begin{tabular}{lccc}
\toprule
Category & User Dissonance & Product Dissonance & Sentiment Dissonance \\
\midrule
Baby Products           & 0.3054 & 0.4733 & 0.1964 \\
Beauty \& Personal Care & 0.3265 & 0.4866 & 0.2162 \\
Pet Supplies            & 0.3114 & 0.4946 & 0.2271 \\
Sports \& Outdoors      & 0.3147 & 0.4949 & 0.1868 \\
Toys \& Games           & 0.3358 & 0.4896 & 0.1898 \\
All Beauty              & 0.2385 & 0.2972 & 0.1534 \\
Arts, Crafts \& Sewing  & 0.3439 & 0.4965 & 0.2001 \\
\bottomrule
\end{tabular}
\end{table*}

\subsection{Interpretation}

All Beauty is comparatively easy for this task: it has stronger text overlap and lower dissonance than the more heterogeneous domains. Categories such as Sports \& Outdoors and Toys \& Games have better rating predictability, but weaker textual alignment. This suggests that some categories support clearer consensus over rating while still permitting more varied narrative expression.

\subsection{Additional Model Comparison Tables}

For completeness, we report the full text-only model comparison tables here rather than in the main paper body.

\begin{table*}[t]
\centering
\caption{Micro metrics for the text-only model comparison on the combined All Beauty setting.}
\label{tab:model_comp_micro}
\footnotesize
\setlength{\tabcolsep}{4pt}
\begin{tabular}{llcccc}
\toprule
Task & Metric & LLaMA-3.1 & GPT-5 & Qwen3-VL & Claude-4.5 \\
\midrule
\multirow{4}{*}{Text}
  & ROUGE-L\,($\uparrow$)     & \textbf{0.1284} & 0.1226 & 0.1258 & 0.1245 \\
  & BLEU\,($\uparrow$)        & \textbf{0.0197} & 0.0113 & 0.0160 & 0.0150 \\
  & METEOR\,($\uparrow$)      & \textbf{0.2072} & 0.1841 & 0.1867 & 0.1972 \\
  & BERT-F1\,($\uparrow$)     & 0.7494 & 0.7474 & 0.7549 & \textbf{0.7569} \\
\addlinespace
\multirow{3}{*}{Title}
  & ROUGE-L\,($\uparrow$)     & 0.0467 & 0.0515 & 0.0591 & \textbf{0.0649} \\
  & BERT-F1\,($\uparrow$)     & 0.7044 & 0.7093 & 0.7148 & \textbf{0.7186} \\
  & Consistency\,($\uparrow$) & 0.4038 & \textbf{0.4793} & 0.4519 & 0.4366 \\
\addlinespace
\multirow{3}{*}{Rating}
  & Accuracy\,($\uparrow$)    & 0.3792 & \textbf{0.5625} & 0.5500 & 0.4750 \\
  & MAE\,($\downarrow$)       & 0.7146 & \textbf{0.6792} & 0.7042 & \textbf{0.6792} \\
  & RMSE\,($\downarrow$)      & \textbf{1.0969} & 1.1850 & 1.2222 & 1.1162 \\
\bottomrule
\end{tabular}
\end{table*}

\begin{table*}[t]
\centering
\caption{Dissonance metrics for the text-only model comparison. Lower is better.}
\label{tab:model_comp_macro}
\footnotesize
\setlength{\tabcolsep}{4pt}
\begin{tabular}{lcccc}
\toprule
Metric\,($\downarrow$) & LLaMA-3.1 & GPT-5 & Qwen3-VL & Claude-4.5 \\
\midrule
User Dissonance      & 0.2651 & \textbf{0.2446} & 0.2452 & 0.2843 \\
Product Dissonance   & \textbf{0.3790} & 0.3843 & 0.3819 & 0.3808 \\
Sentiment Dissonance & \textbf{0.1458} & 0.1518 & 0.1663 & 0.1585 \\
\bottomrule
\end{tabular}
\end{table*}

\section{Dissonance and Evaluation Details}
\label{sec:appendix_discriminator}

\subsection{Rationale}

Standard overlap metrics do not tell us whether a generation behaves like the target user or remains close to product consensus. Dissonance Analysis is meant to summarize those higher-level mismatches. These scores are heuristic and should be read as complements to the standard text metrics.

\subsection{Operational Formulas}
\label{sec:appendix_dissonance}

User Dissonance measures deviation from the target user's historical profile:
\begin{equation}
\label{eq:user_dissonance_appendix}
D_{\mathrm{user}} =
w_1 |\Delta \mathrm{style}| +
w_2 |\Delta \mathrm{sentiment}| +
w_3 |\Delta \mathrm{rating}| +
w_4 |\Delta \mathrm{length}|.
\end{equation}

Product Dissonance measures drift away from the product's review cluster:
\begin{equation}
\label{eq:product_dissonance_appendix}
D_{\mathrm{prod}} =
\left\| \mathbf{e}_{\mathrm{pred}} - \mathbf{c}_{\mathrm{cluster}} \right\|_2 +
\mathrm{aspect\ divergence}.
\end{equation}

Sentiment Dissonance measures disagreement among text sentiment, predicted rating, and gold behavior:
\begin{equation}
\label{eq:sentiment_dissonance_appendix}
D_{\mathrm{sent}} =
|\mathrm{sent}_{\mathrm{pred}} - \mathrm{sent}_{\mathrm{gold}}| +
|\mathrm{sent}_{\mathrm{pred}} - \mathrm{rating}_{\mathrm{pred}}| +
|\mathrm{rating}_{\mathrm{pred}} - \mathrm{rating}_{\mathrm{gold}}|.
\end{equation}

\subsection{Style Parameter Scope and Missing Ablation}
\label{sec:appendix_style}

The User Style Parameter intentionally uses a compact 11-feature summary: four length features, four VADER sentiment features, and three writing-style features. This choice favors interpretability and stability under sparse user histories.

The current submission does not include a leave-one-feature-group-out style ablation. In other words, we do not yet isolate how much comes from length features, sentiment features, or punctuation and pronoun features separately. This should be treated as a limitation rather than as evidence that all feature groups contribute equally.

\subsection{Metric-Component Ablation Status}
\label{sec:appendix_metric_ablation}

The current submission also does not include an ablation over the Dissonance metric components themselves. We provide the operational formulas above, but we do not separately test, for example, whether the style term dominates User Dissonance or whether the aspect-divergence term dominates Product Dissonance. This is another limitation of the current evaluation package.

\subsection{Interpretation Limits}

Dissonance scores rely on heuristic proxies. Product Dissonance assumes that the product-neighbor cluster is a reasonable reference point, which may penalize novel but valid observations. User Dissonance assumes that a compact stylometric profile is a useful proxy for stable user behavior. These scores should therefore be interpreted as behavioral summaries, not as exact measurements of truth.

\section{Qualitative Analysis}
\label{sec:qualitative_analysis}

\begin{figure}[t]
    \centering
    \includegraphics[width=0.95\columnwidth]{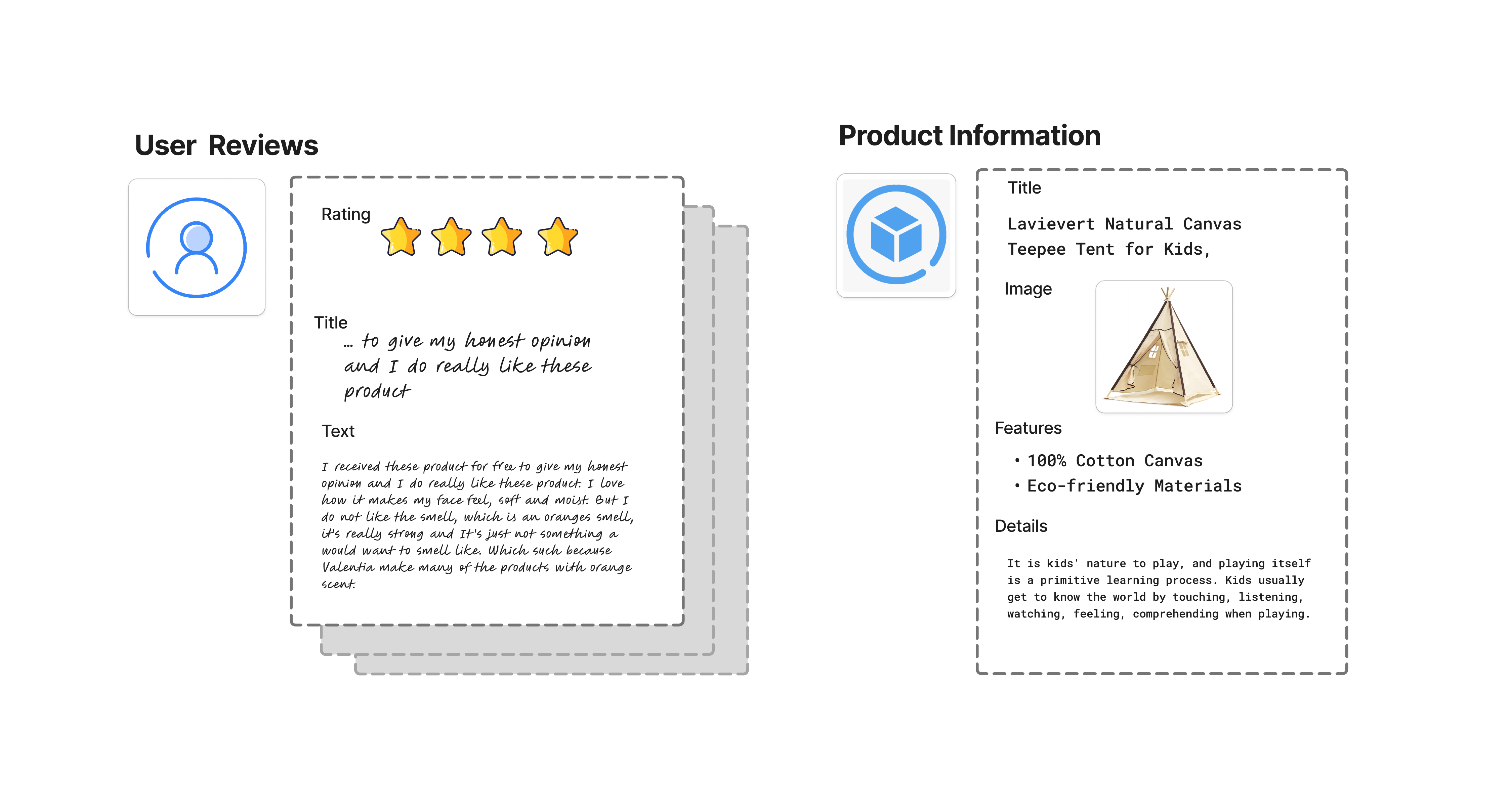}
    \caption{Illustration of the input structure. The user profile provides stylistic evidence, while product information provides grounding context.}
    \label{fig:qualitative_visual}
\end{figure}

To illustrate the role of visual context, Figure~\ref{fig:qualitative_visual} shows the input structure used by the system, and Table~\ref{tab:qualitative_examples} gives two representative examples where visual grounding affects the final output.

\begin{table*}[t]
\centering
\small
\renewcommand{\arraystretch}{1.3}
\setlength{\tabcolsep}{4pt}
\begin{tabular}{p{1.8cm} p{1.4cm} p{9.7cm}}
\toprule
Model & Rating & Generated review and short analysis \\
\midrule
\multicolumn{3}{l}{Example A: correcting sentiment drift} \\
\rowcolor{goldbg}
Gold & 4.0 & Title: Rich shades of red. The gold review is positive overall, with mild complaints about wear and lip dryness. \\
\rowcolor{textbg}
Text-only & 2.0 & The text-only system hallucinates a packaging or shipping failure and drifts negative. This appears to come from over-reliance on user-history complaints. \\
\rowcolor{multibg}
Multimodal & 4.0 & The multimodal output aligns with the product context and keeps the review near the correct positive rating. \\
\midrule
\multicolumn{3}{l}{Example B: detecting negative quality cues} \\
\rowcolor{goldbg}
Gold & 2.0 & Title: I loved the idea of this organic nail polish remover. The gold review is clearly negative despite an initially positive setup. \\
\rowcolor{textbg}
Text-only & 4.0 & The text-only system defaults to a generic positive review and misses the product-specific failure case. \\
\rowcolor{multibg}
Multimodal & 1.0 & The multimodal output is harsher than gold, but it captures the negative product experience much better than the text-only baseline. \\
\bottomrule
\end{tabular}
\caption{Short qualitative examples comparing text-only and multimodal behavior.}
\label{tab:qualitative_examples}
\end{table*}

\end{document}